\documentclass[a4paper]{article}
\usepackage[pdftex]{graphicx}
\usepackage{amsmath,amssymb,amsthm}
\usepackage{dsfont,array}
\usepackage{hyperref}

\usepackage{triotex}
\usepackage{mathpartir}
\usepackage{listings}
\usepackage{xcolor}
\theoremstyle{plain}
\newtheorem{theorem}{Theorem}
\newtheorem{proposition}[theorem]{Proposition}

\theoremstyle{definition}

\newcommand{\qcal}{\mathcal{Q}}

\newcommand{\cat}{\star}
\newcommand{\catl}{\circ}

\newcommand{\scope}{\;\bullet\;}

\newcommand{\first}{\mathsf{fst}}
\newcommand{\last}{\mathsf{lst}}

\newcommand{\decp}{\mathcal{D}_{\mathsf{seq}(\integers)}}
\newcommand{\catsig}{\Sigma_{\mathsf{cat}}}
\newcommand{\pressig}{\Sigma_{\integers}}
\newcommand{\catth}{\mathcal{T}_{\mathsf{cat}}}
\newcommand{\seqth}{\mathcal{T}_{\mathsf{seq}(\integers)}}
\newcommand{\seqsig}{\Sigma_{\mathsf{seq}(\integers)}}
\newcommand{\rcal}{\mathcal{R}}
\newcommand{\reg}{\mathrm{R}}

\newcommand{\uu}{\mathsf{u}}
\newcommand{\vv}{\mathsf{v}}
\newcommand{\sorted}{\mathrm{sorted}}
\newcommand{\Result}{\mathbf{Result}}
\newcommand{\old}{\mathbf{old}}
\newcommand{\rev}{^\mathsf{R}}

\newcommand{\adp}{\mathcal{ADP}}
\newcommand{\sil}{\textbf{SIL}}
\newcommand{\lia}{\textbf{LIA}}

\title{\textsc{What's Decidable About Sequences?}}

\author{Carlo A. Furia}

\date{January 2010}

\begin{document}

\maketitle

\vspace{3cm}
\begin{abstract}
We present a first-order theory of sequences with integer elements, Presburger arithmetic, and regular constraints, which can model significant properties of data structures such as arrays and lists.
We give a decision procedure for the quantifier-free fragment, based on an encoding into the first-order theory of concatenation; the procedure has PSPACE complexity.
The quantifier-free fragment of the theory of sequences can express properties such as sortedness and injectivity, as well as Boolean combinations of periodic and arithmetic facts relating the elements of the sequence and their positions (e.g., ``for all even $i$'s, the element at position $i$ has value $i+3$ or $2i$'').
The resulting expressive power is orthogonal to that of the most expressive decidable logics for arrays.
Some examples demonstrate that the fragment is also suitable to reason about sequence-manipulating programs within the standard framework of axiomatic semantics.
\end{abstract}

\pagebreak

\tableofcontents

\pagebreak

\section{Introduction}
\label{sec:introduction}
Verification is undecidable already for simple programs, but modern programming languages support a variety of sophisticated features that make it all the more complicated.
These advanced features --- such as arrays, pointers, dynamic allocation of resources, and object-oriented abstract data types --- are needed because they raise the level of abstraction thus making programmers more productive and programs less buggy.
Verification techniques have also progressed rapidly over the years, in an attempt to keep the pace with the development of programming languages.

Automated verification requires expressive program logics and powerful decision procedures.
In response to the evolution of modern programming languages, new decidable program logic fragments and combination techniques for different fragments have mushroomed especially in recent years.
Many of the most successful contributions have focused on verifying relatively restricted aspects of a program's behavior, for example by decoupling pointer structure and functional properties in the formal analysis of a dynamic data structure.
This narrowing choice, partly deliberate and partly required by the formidable difficulty of the various problems, is effective because different aspects are often sufficiently decoupled so that each of them can be analyzed in isolation with the most appropriate, specific technique.

This paper contributes to the growing repertory of special program logics by exploring the decidability of properties of \emph{sequences of elements} of homogeneous type.
These can abstract fundamental features of several data structures: arrays imprimis, but also the sequence of values stored in a dynamically allocated list, or the content of a stack or a queue.

We take a new angle on reasoning about sequences, based on the \emph{theory of concatenation}: a first-order theory where variables are interpreted as words (or sequences) over a finite alphabet and can be composed by concatenating them.
Makanin's algorithm for solving word equations \cite{Makanin-algo} implies the decidability of the quantifier-free fragment of the theory of concatenation.
Based on this, we introduce a first-order \emph{theory of sequences} $\seqth$ whose elements are integers.
Section \ref{sec:decision} presents a decision procedure for the quantifier-free fragment of $\seqth$, which encodes the validity problem into the quantifier-free theory of concatenation.
The decision procedure is in PSPACE; it is known, however, that Makanin's algorithm is reasonably efficient in practice \cite{AP-implementation}.

The theory of sequences $\seqth$ allows concatenating sequences to build new ones, and it includes Presburger arithmetic over elements of a sequence.
On the other hand, it forbids explicit indexed access to elements, which differentiates it from the theory of arrays and extensions thereof (see Section \ref{sec:related}).
The resulting quantifier-free fragment has significant expressiveness, in spite of its limitations in representing subsequences of variable length.
In particular, we show some interesting properties that are inexpressible in powerful decidable array logics (such as those in \cite{BMS06-vmcai,GNRZ07-amai,HIV08-fossacs,HIV08-lpar}) but are expressible in our theory of sequences.
Conversely, there exist decidable properties of extensions of the theory of arrays that are inexpressible in $\seqth$.
These results support our claim that the theory of sequences provides a fresh angle on reasoning about sequences, orthogonal to most approaches that model sequences as arrays.

In order to better assess the limits of our theory of sequences, we also prove that several natural extensions of the quantifier-free fragment of $\seqth$ are undecidable.
Finally, we demonstrate reasoning about sequence-manipulating programs with annotations written in the quantifier-free fragment of $\seqth$.
A couple of examples in Section \ref{sec:programs} illustrate the usage of $\seqth$ formulas with the standard machinery of axiomatic semantics and backward reasoning.

\paragraph{Paper outline.}
Section \ref{sec:concatenation} presents the theory of concatenation and summarizes a few decidability and undecidability results about it.
Section \ref{sec:sequences} introduces our theory of integer sequences $\seqth$, demonstrates its expressiveness, provides a decision procedure for its quantifier-free fragment, and shows undecidable extensions of the theory.
Section \ref{sec:programs} illustrates how to use the theory $\seqth$ to reason about programs in the standard axiomatic semantics framework.
Finally, Section \ref{sec:related} reviews related work and Section \ref{sec:conclusion} concludes by outlining future work.

\section{The Theory of Concatenation}
\label{sec:concatenation}
This section introduces some basic notation (Section \ref{sec:preliminaries}) and summarizes some results about the first-order theory of concatenation (Section \ref{sec:dec-concatenation}) that we will use in the remainder of the paper.

In the rest of the paper, we assume familiarity with the standard syntax and terminology of first-order theories (e.g., \cite{BM-book}); in particular, we assume the standard abbreviations and symbols of first-order theories with the following operator precedence $\neg, \wedge, \vee, \Rightarrow, \Leftrightarrow, \forall \text{ and } \exists$.

$FV(\phi)$ denotes the set of free variables of a formula $\phi$.
With standard terminology, a formula $\phi$ is a \emph{sentence} iff it is \emph{closed} iff $FV(\phi) = \emptyset$.
Given a regular expression $\qcal$ over $\{\exists, \forall\}$, the $\qcal$-fragment of a first-order theory is the set of all formulas of the theory in the form $\qcal \scope \psi$, where $\psi$ is quantifier-free.
The universal and existential fragments are synonyms for the $\forall^*$- and $\exists^*$-fragment respectively.
A fragment is decidable iff the validity problem is decidable for its sentences.
It is customary to define the validity and satisfiability problems for a quantifier-free formula $\psi$ as follows: $\psi$ is valid iff the universal closure of $\psi$ is valid, and $\psi$ is satisfiable iff the existential closure of $\psi$ is valid.
As a consequence of this definition, the decidability of a quantifier-free fragment whose formulas are closed under negation is tantamount to the decidability of the universal or existential fragments.
Correspondingly, in the paper we will allow some freedom in picking the terminology that is most appropriate to the context.

\subsection{Sequences and Concatenation}
\label{sec:preliminaries}
$\integers$ denotes the set of integer numbers and $\naturals$ denotes the set of nonnegative integers.

Given a set $A = \{a, b, c, \ldots \}$ of constants, a \emph{sequence} over $A$ is any word $v = v(1) v(2) \cdots v(n)$ for some $n \in \naturals$ where $v(i) \in A$ for all $1 \leq i \leq n$.
The symbol $\epsilon$ denotes the \emph{empty sequence}, for which $n = 0$.
$|v| = n$ denotes the \emph{length} of $v$.
$A^*$ denotes the set of all finite sequences over $A$ including $\epsilon \not\in A$.
It is also convenient to introduce the shorthand $v(k_1, k_2)$ with $k_1, k_2 \in \integers$ to describe \emph{subsequences} of a given sequence $v$; it is defined as follows.
\begin{equation*}
v(k_1, k_2) \triangleq
\begin{cases}
v(k_1) v(k_1 + 1) \cdots v(k_2)				&	1 \leq k_1 \leq k_2 \leq |v|			\\
v(k_1, |v|+k_2)								&   k_1 -|v| \leq k_2 < 1 \leq k_1			\\
v(|v|+k_1, |v|+k_2)							&   1 -|v| \leq k_1 \leq k_2 < 1			\\
\epsilon									&   \text{otherwise}
\end{cases}
\end{equation*}
For two sequences $v_1, v_2 \in A^*$, $v_1 \cat v_2$ denotes their \emph{concatenation}: the sequence $v_1(1) \cdots v_1(|v_1|)v_2(1) \cdots v_2(|v_2|)$.
We will drop the concatenation symbol whenever unambiguous.

The structure $\langle A^*, \cat, \epsilon \rangle$ is also referred to as the \emph{free monoid} with generators in $A$ and neutral element $\epsilon$.
The size $|A|$ is called \emph{rank} of the free monoid and it can be finite or infinite.

\subsection{Decidability in the Theory of Concatenation}
\label{sec:dec-concatenation}

\subsubsection{Syntax and Semantics}
The theory of concatenation is the first-order theory $\catth$ with signature
\begin{equation*}
\catsig \quad\triangleq\quad \{\doteq, \catl, \rcal\}
\end{equation*}
where $\doteq$ is the equality predicate,\footnote{We use the symbol $\doteq$ to distinguish it from the standard arithmetic equality symbol $=$ used later in the paper.} $\catl$ is the binary concatenation function and $\rcal \triangleq \{\reg_1, \reg_2, \ldots\}$ is a set of unary (monadic) predicate symbols called \emph{regularity constraints}.
We sometimes write $R_i(x)$ as $x \in R_i$ and $\alpha \not\doteq \beta$ abbreviates $\neg (\alpha \doteq \beta)$.

An interpretation of a formula in the theory of concatenation is a structure $\langle A^*, \cat, \epsilon, \underline{\rcal}, ev \rangle$ where $\langle A^*, \cat, \epsilon \rangle$ is a free monoid, $\underline{\rcal} = \{\underline{\reg}_1, \underline{\reg}_2, \ldots\}$ is a collection of regular subsets of $A^*$, and $ev$ is a mapping from variables to values in $A^*$.
The satisfaction relation $\langle A^*, \cat, \epsilon, \underline{\rcal}, ev \rangle \models \phi$ for formulas in $\catth$ is defined in a standard fashion with the following assumptions.
\begin{itemize}
\item any variable $x$ takes the value $ev(x) \in A^*$;
\item the concatenation $x \catl y$ of two variables $x,y$ takes the value $ev(x) \cat ev(y)$;
\item for each $\reg_i \in \rcal$, the corresponding $\underline{\reg}_i \in \underline{\rcal}$ defines the set of sequences $x \in \underline{\reg}_i$ for which $\reg_i(x)$ holds (this also subsumes the usage of constants).
\end{itemize}

\subsubsection{(Un)Decidable Fragments}
The following propositions summarize some decidability results about fragments of the theory of concatenation; they all are known results, or corollaries of them.
The standard presentation of these results focuses on solving equations over sequences with free variables and, correspondingly, on existential fragments of the equational theory.
On the contrary, in this paper we will mostly focus on the universal fragment, given its aptness for annotating sequence-manipulating programs (see Section \ref{sec:programs}).
It is straightforward, however, to rephrase the results in terms of the dual existential fragments, given the availability of negation in the language.

\begin{proposition}[Decidability \cite{Makanin-algo,Makanin-survey,Pla04}] \label{prop:decidable}
The universal and existential fragments of the theory of concatenation over free monoids with finite rank are decidable in PSPACE.
\end{proposition}
\begin{proof}
Decidability is a consequence of Makanin's seminal result on word equations \cite{Makanin-algo} and its extensions to the full existential (and universal) fragments \cite{BS-coding,Makanin-survey}.
PSPACE complexity is a consequence of Plandowski's recent results \cite{Pla04,Pla06} and the fact that transforming first-order formulas into a single word equation introduces only a polynomial blow-up.

The only catch is that the standard presentation assumes formulas in the canonical form $\forall x_1 \in \reg_1, x_2\in \reg_2, \ldots, x_v\in \reg_v \scope \rho$ where regularity constraints do not appear in $\rho$.
This is, however, without loss of generality as we can put any universal formula $\forall \overline{x} \scope \psi$ in canonical form: first rewrite $\psi$ into
\begin{equation*}
\bigwedge_{\substack{1 \leq i \leq |\overline{x}|  \\  1 \leq j \leq |\rcal|}}  (x_i \doteq h^+_j \vee x_i \doteq h^-_j)
	\Rightarrow \psi
\end{equation*}
for fresh $h^+_j \in \reg_j$ and $h^-_j \in A^* \setminus \reg_j$.
Then, put $\psi$ in negated normal form and eliminate occurrences of regularity predicates by applying exhaustively the rules:
\begin{equation*}
\inferrule{\psi[\reg_m(x_n)]}{\psi[x_n \doteq h^+_m]}
\qquad
\inferrule{\psi[\neg \reg_m(x_n)]}{\psi[x_n \doteq h^-_m]}
\end{equation*}
It is not difficult to see that this transformation preserves satisfiability and introduces a blow-up which is quadratic at most.
\end{proof}

\begin{proposition}[Undecidability] \label{prop:undecidable}
\begin{itemize}
\item\cite{Dur95}
The $\forall^*\exists^*$ and $\exists^*\forall^*$ fragments of the theory of concatenation are undecidable; in particular the $\forall \exists^3$-fragment is undecidable already for negation-free formulas.

\item\cite{BS-definability}
The existential and universal fragments of the extension of the theory of concatenation over the free monoid $\{a,b\}^*$ with: (1) two length functions $|x|_a \triangleq \{ y \in a^* \mid y \text{ has the same number of $a$'s as } x\}$ and $|x|_b \triangleq \{ y \in b^* \mid y \text{ has the same number of $b$'s as } x\}$; or (2) the function $Sp(x) \triangleq |x|_a \cat |x|_b$ are undecidable.
\end{itemize}
\end{proposition}

A set of sequences $S \subseteq A^*$ is \emph{universally (resp.\ existentially) definable from concatenation} iff there is a universal (resp.\ existential) formula $\varphi[x]$ with $FV(\varphi[x]) = \{ x \}$ such that $S = \{ y \in A^* \mid \varphi[y] \}$.
\begin{proposition}[Definability \cite{BS-definability}] \label{prop:definable}
\begin{itemize}
\item The set $S^= \triangleq \{ a^n b^n \mid n \in \naturals \}$ is neither universally nor existentially definable from concatenation.
\item The equal length predicate $Elg(x,y) \triangleq |x| = |y|$ is not definable in the existential and universal fragments of concatenation.
\end{itemize}
\end{proposition}
\begin{proof}
B\"uchi and Senger prove in \cite[Corollary 3]{BS-definability} that $S^=$ is not existentially definable.
A very similar argument shows that $S^< \triangleq \{ a^m b^n \mid 0 \leq m < n \}$ is also not existentially definable (using the terminology of \cite[Corollary 3]{BS-definability}, the spacers of $a^{i-1}b^i$ relative to the atom $a$ are $\langle a, ab^i\rangle$ and there are only $i-1$ words in $S^<$ with these spacers).
The existential non-definability of $S^<$ entails the existential non-definability of the set $S^{\neq} \triangleq \{a^n b^m \mid 0 \leq n \neq m\}$ by contradiction as follows.
Assume that $S^{\neq}$ were existentially definable; then $x \in S^<$ could be defined as $x \in S^{\neq} \wedge \exists u, v, p (u \in a^* \wedge v \in b^* \wedge p \in a^+ \wedge upv \not\in S^{\neq})$ (that is, $|u| + |p| = |v|$), a contradiction.
Finally, $S^=$ is universally definable from concatenation iff $S^{\neq}$ is existentially definable from concatenation.
In fact, the complement set $\{a,b\} \setminus S^=$ is $S^\sim \cup S^{\neq}$ with $S^\sim \triangleq \{a,b\}^*b\{a,b\}^*a\{a,b\}^*$ clearly existentially and universally definable from concatenation.
This concludes the proof of the first part of the proposition.

The second part is proved in \cite[Theorem 1]{BS-definability} for the existential fragment and it is straightforward to adapt that proof to universal definability.
\end{proof}

It is currently unknown whether the extension of the existential or universal fragment of concatenation with $Elg$ is decidable, while allowing membership constraints over deterministic context-free language gives an undecidable theory \cite{Makanin-survey}.

\section{A Theory of Sequences}
\label{sec:sequences}
This section introduces a first-order theory of sequences (Section \ref{sec:seq-theory-def}) with arithmetic, gives a decision procedure for its universal fragment (Section \ref{sec:decision}), and shows that ``natural'' larger fragments are undecidable (Section \ref{sec:extensions}).

\subsection{A Theory of Integer Sequences} \label{sec:seq-theory-def}
We present an arithmetic theory of sequences whose elements are integers.
It would be possible to make the theory parametric with respect to the element type.
Focusing on integers, however, makes the presentation clearer and more concrete, with minimal loss of generality as one can introduce any theory definable in the integer arithmetic fragment.

\subsubsection{Syntax and Semantics}

\paragraph{Syntax.}
Properties of integers are expressed in Presburger arithmetic whose signature is:
\begin{equation*}
\pressig \quad\triangleq\quad \{0, 1, +, -, =, <\}
\end{equation*}
Then, our theory $\seqth$ of sequences with integer values has signature
\begin{equation*}
\seqsig \quad\triangleq\quad \catsig \cup \pressig
\end{equation*}
Operator precedence is: $\catl; + \text{ and } -; \doteq, = \text{ and } <$ followed by logic connectives and quantifiers with the previously defined precedence.

We will generally consider formulas in prenex normal form
\begin{equation*}
\qcal \scope \psi
\end{equation*}
where $\qcal$ is a quantifier prefix and $\psi$ is quantifier-free written in the grammar:
\begin{equation*}
\begin{split}
seq 	\quad & ::= \quad var  \mid  int  \mid  seq \catl seq  \\
int 	\quad & ::= \quad 0  \mid  1  \mid  seq  \mid  int + int  \mid  int - int \\
fmla 	\quad & ::= \quad seq \doteq seq  \mid  R(seq)  \mid  int = int  \mid  int < int    \\
              &   \qquad \mid \neg fmla  \mid  fmla \vee fmla  \mid fmla \wedge fmla \mid fmla \Rightarrow fmla
\end{split}
\end{equation*}
with $var$ ranging over variable names.

\paragraph{Semantics.}
An interpretation of a sentence of $\seqth$ is a structure $\langle \integers^*, \cat, \epsilon, \underline{\rcal},\linebreak ev\rangle$ with the following assumptions.\footnote{The presentation of the semantics of the theory is informal and implicit for brevity.}
\begin{itemize}
\item $\langle \integers^*, \cat, \epsilon, \underline{\rcal}, ev \rangle$ have the same meaning as in the theory of concatenation.
\item As far as arithmetic is concerned: 
	\begin{itemize}
	\item The interpretation of a sequence $v_1 v_2 \cdots \in \integers^*$ of integers is the first integer in the sequence $v_1$, with the convention that the interpretation of the empty sequence is $0$.
	\item Conversely, the interpretation of an integer value $v \in \integers$ is the singleton sequence $v$.
	\item Addition, subtraction, equality, and less than are interpreted accordingly.
	\end{itemize}
\end{itemize}
The satisfaction relation is then defined in a standard fashion.

\paragraph{Shorthands.}
We introduce several shorthands to simplify the writing of complex formulas.
\begin{itemize}
\item A symbol for every constant $k \in \integers$, defined as obvious.
\item $\alpha \neq \beta$, $\alpha \leq \beta$, $\alpha \geq \beta$, and $\alpha > \beta$ defined respectively as $\neg (\alpha = \beta)$, $\alpha < \beta \vee \alpha = \beta$, $\neg (\alpha < \beta)$, and $\alpha \geq \beta \wedge \alpha \neq \beta$.
\item Shorthands such as $\alpha \leq \beta < \gamma$ or $\beta \in [\alpha, \gamma)$ for $\alpha \leq \beta \wedge \beta < \gamma$.
\item Bounded length predicates such as $|x| < k$ for a variable $x$ and a constant $k \in \integers$ abbreviating $\widehat{\reg}^{< k}(x)$ with $\widehat{\reg}^{< k}$ a regular constraint interpreted as $\{\epsilon\} \cup \bigcup_{0 < i < k} \integers^i$.
The definition of derived expressions such as $k_1 \leq |x| < k_2$ is also as obvious.
\item Subsequence functions such as $x(k_1, k_2)$ for a variable $x$ and two constants $k_1, k_2 \in \integers$ with the intended semantics (see Section \ref{sec:preliminaries}).
	We define these functions in the theory $\seqth$ by the following rewriting rules, defined on formulas in prenex normal form with quantifier prefix $\qcal$:
	\begin{scriptsize}
	\begin{equation*}
		\begin{array}{c}
		\qcal \scope \psi[x(k_1, k_2)] \\
		\hline \\
		\qcal \forall u,v,w \scope
			\left( \begin{array}{l}
			\kappa_1 \wedge x \doteq uvw \wedge |u| = k_1 - 1 \wedge |v| = k_2 - k_1 + 1 \\
			\vee\quad   \kappa_2 \wedge x \doteq uvw \wedge |u| = k_1 - 1 \wedge |w| = - k_2  \\
			\vee\quad   \kappa_3 \wedge x \doteq uvw \wedge |v| = -k_1 + k_2 +1 \wedge |w| = - k_2 \\
			\vee\quad   \neg (\kappa_1 \vee \kappa_2 \vee \kappa_3) \wedge u \doteq v \doteq w \doteq \epsilon 
			\end{array} \right)
		\Rightarrow \psi[v]
		\end{array}
	\end{equation*}
	\end{scriptsize}
where: 
\begin{align*}
\kappa_1 & \quad \triangleq \quad		1 \leq k_1 \leq k_2 \leq |x|  		\\
\kappa_2 & \quad \triangleq \quad 		k_1 -|x| \leq k_2 < 1 \leq k_1		\\
\kappa_3 & \quad \triangleq \quad 		1 -|x| \leq k_1 \leq k_2 < 1
\end{align*}
\item $\first(x)$ and $\last(x)$ for the first $x(1,1)$ and last element $x(0,0)$ of $x$, respectively.
\end{itemize}

\subsubsection{Examples}
\label{sec:examples}
A few examples demonstrate the expressiveness of the universal fragment of $\seqth$ to specify properties of sequences.

\begin{enumerate}
\item \textbf{Equality}: sequences $\uu$ and $\vv$ are equal.
		\begin{equation}
		\uu \doteq \vv
		\end{equation}

\item \textbf{Bounded equality}: sequences $\uu$ and $\vv$ are equal in the \emph{constant} interval $[l, u]$ for $l, u \in \integers$.
		\begin{equation}
		\uu(l,u) \doteq \vv(l,u)
		\end{equation}

\item \textbf{Boundedness}: no element in sequence $\uu$ is greater than value $\vv$.
		\begin{equation} \label{eq:boundedness}
		\forall h,t \scope \uu \doteq ht \;\Rightarrow\; t \leq \vv
		\end{equation}

\item \textbf{Sortedness}: sequence $\uu$ is sorted (strictly increasing).
		\begin{equation} \label{eq:strictly-increasing}
		\forall h,m,t \scope \uu \doteq hmt \wedge |m| = 1 \wedge |t| > 0 \;\Rightarrow\; m < t
		\end{equation}

\item \textbf{Injectivity}: $\uu$ has no repeated elements.
		\begin{equation} \label{eq:injective}
		\forall h, v_1, m, v_2, t \scope   \uu \doteq h v_1 m v_2 t \wedge |v_1| = 1 \wedge |v_2| =1
					\Rightarrow v_1 \neq v_2
		\end{equation}

\item \textbf{Partitioning}: sequence $\uu$ is partitioned at \emph{constant} position $k > 0$.
		\begin{equation}
		\forall h_1,t_1,h_2,t_2 \scope  \left( \begin{array}{l}
										\uu(1,k) \doteq h_1t_1  \\
										\wedge\  \uu(k+1, 0) \doteq h_2t_2  \\
										\wedge\  |t_1| > 0 \wedge |t_2| > 0
										\end{array} \right)
											\Rightarrow   t_1 < t_2
		\end{equation}

\item \textbf{Membership}: \emph{constant} element $k \in \integers$ occurs in sequence $\uu$.
		\begin{equation}
		\uu \in (\integers^* k \integers^*)
		\end{equation}

\item \textbf{Non-membership}: no element in sequence $\uu$ has value $\vv$.
		\begin{equation}
		\forall h,t \scope \uu \doteq ht \wedge |t| > 0 \;\Rightarrow\; t \neq \vv
		\end{equation}

\item \textbf{Periodicity}: in non-empty sequence $\uu$, elements on even positions have value $0$ and elements on odd positions have value $1$ (notice that $\last(h) = 0$ if $h$ is empty).
		\begin{equation} \label{eq:periodicity}
		\forall h,t \scope   	\uu \doteq ht \wedge |t| > 0
								\Rightarrow
								\left( \begin{array}{c}
								\last(h) = 1 \\
								\Rightarrow t = 0
								\end{array} \right)
									\wedge
								\left( \begin{array}{c}
								\last(h) = 0 \\
								\Rightarrow t = 1
								\end{array} \right)
		\end{equation}

\item \textbf{Comparison} between indices and values: for every index $i$, element at position $i$ in the non-empty sequence $\uu$ has  value $i + 3$.
		\begin{equation} \label{eq:comparison}
		\uu = 1 + 3 \wedge
		\forall h,t,v \scope   \uu \doteq ht \wedge |h| > 0 \wedge |t| > 0 \wedge \last(h) = v	\Rightarrow	t = v + 1
		\end{equation}	

\item \textbf{Disjunction of value constraints}: for every pair of positions $i < j$ in the sequence $\uu$, either $\uu(i,i) \leq \uu(j,j)$ or $\uu(i,i) \geq 2 \uu(j,j)$.
		\begin{equation} \label{eq:disjunction}
		\forall h, v_1, m, v_2, t \scope   \uu \doteq h v_1 m v_2 t \wedge |v_1| > 0 \wedge |v_2| > 0
					\Rightarrow v_1 \leq v_2 \vee v_1 \geq v_2 + v_2
		\end{equation}

\end{enumerate}

\paragraph{Comparison with theories of arrays.}
Properties such as strict sortedness (\ref{eq:strictly-increasing}), periodicity (\ref{eq:periodicity}), and comparisons between indices and values (\ref{eq:comparison}) are inexpressible in the array logic of Bradley et al.~\cite{BMS06-vmcai}.
The latter is inexpressible also in the logic of Ghilardi et al.~\cite{GNRZ07-amai} because Presburger arithmetic is restricted to indices.
Properties such as (\ref{eq:disjunction}) are inexpressible both in the SIL array logic of \cite{HIV08-lpar} --- because quantification on multiple array indices is disallowed --- and in the related LIA logic of \cite{HIV08-fossacs} --- because disjunctions of comparisons of array elements are disallowed.
Extensions of each of these logics to accommodate the required features would be undecidable.

Conversely, properties such as \emph{permutation}, bounded equality \emph{for an interval specified by indices}, length constraints \emph{for a variable value}, membership \emph{for a variable value}, and the \emph{subsequence} relation, are inexpressible in the universal fragment of $\seqth$.
Notice that membership and the subsequence relation are expressible in the dual existential fragment of $\seqth$, while the other properties seem to entail undecidability of the corresponding $\seqth$ fragment (see Section \ref{sec:extensions}).
Bounded equality, length constraints, and membership, on the other hand, are expressible in all the logics of \cite{BMS06-vmcai,GNRZ07-amai,HIV08-lpar,HIV08-fossacs}, and \cite{GNRZ07-amai} outlines a decidable extension which supports the subsequence relation (see Section \ref{sec:related}).

\subsection{Deciding Properties of Integer Sequences} \label{sec:decision}
This section presents a decision procedure $\decp$ for the universal fragment of $\seqth$.
The procedure transforms any universal $\seqth$ formula into an equi-satisfiable universal formula in the theory of concatenation over the free monoid $\{a,b,c,d\}^*$.
The basic idea is to encode integers as sequences over the four symbols $\{a,b,c,d\}$: the sequence $acb^{k_1}a$ encodes a nonnegative integer $k_1$, while the sequence $adb^{-k_2}a$ encodes a negative integer $k_2$.
Suitable rewrite rules encode all quantifier-free Presburger arithmetic in accordance with this convention.
The next subsection \ref{sec:dec-proc} outlines the decision procedure $\decp$, while subsection \ref{sec:correctness} illustrates its correctness and discusses its complexity.

\subsubsection{$\decp$: A Decision Procedure for $\seqth$}
\label{sec:dec-proc}

Consider a universal formula of $\seqth$ in prenex normal form:
\begin{equation}
\forall x_1, \ldots, x_v \scope \psi 
\label{eq:prenex}
\end{equation}
where $\psi$ is quantifier-free.
Modify (\ref{eq:prenex}) by application of the following steps.

\begin{enumerate}
\item Introduce fresh variables to normalize formulas into the following form:
\begin{equation*}
\begin{split}
fmla 	\quad & ::=  \quad var \doteq var \mid var \doteq var \catl var  \mid  R(var) \mid var = 0 \mid var = 1 \\
			  & \qquad \mid  var = var \mid  var = var + var  \mid  var = var - var \mid  var < var \\
			  & \qquad \mid \neg fmla  \mid  fmla \vee fmla  \mid fmla \wedge fmla \mid fmla \Rightarrow fmla
\end{split}
\end{equation*}
Clearly, we can achieve this by applying exhaustively rewrite rules that operate on $\psi$ such as:
\begin{equation*}
\inferrule{\psi[x \catl y]}{e \doteq x \catl y  \Rightarrow \psi[e]}    \qquad
\inferrule{\psi[x + y]}{f = x + y  \Rightarrow \psi[f]}
\end{equation*}
for fresh variables $e,f$.


\item For each variable $x_i \in FV(\psi) = \{x_1, \ldots, x_v\}$, introduce the fresh variables $h_i, t_i, s_i, m_i$ (for head, tail, sign, modulus) and rewrite $\psi$ as:
\begin{equation*}
	\bigwedge_{1 \leq i \leq v}
	\left( \left( \begin{array}{l}
	x_i \doteq h_i t_i \\
		\wedge\  h_i \doteq a s_i m_i a \\
		\wedge\  s_i \in \{c, d\} \\
		\wedge\  m_i \in b^* \\
		\wedge\  t_i \in (acb^*a \cup adb^+a)^*
	\end{array} \right)
		\vee
	\left( \begin{array}{l}
	x_i \doteq \epsilon \\
		\wedge\  h_i \doteq a s_i m_i a  \\
		\wedge\  s_i \doteq c \\
		\wedge\  m_i \doteq \epsilon \\
		\wedge\  t_i \doteq \epsilon
	\end{array} \right) \right)
			\Rightarrow  \psi
\label{eq:arithmetic}
\end{equation*}

\item Apply the following rule exhaustively to remove arithmetic equalities: \\
\begin{equation*}
\inferrule{\psi[x_i = x_j]}{\psi[h_i \doteq h_j]}
\qquad
\inferrule{\psi[x_i = 0]}{\psi[h_i \in 0]}
\qquad
\inferrule{\psi[x_i = 1]}{\psi[h_i \in 1]}
\end{equation*}

\item Apply the following rule exhaustively to remove differences: \\
\begin{equation*}
\inferrule{\psi[x_k = x_i - x_j]}{\psi[x_i = x_k + x_j]}
\end{equation*}

\item Apply the following rule exhaustively to remove comparisons:
\begin{equation*}
	\begin{array}{c}
	\psi[x_i < x_j] \\
	\hline \\
	\left(\begin{array}{l}
		m_i \doteq m_j  \\
		\vee\; m_i \doteq m_j p \\
		\vee\; m_j \doteq m_i p
	\end{array} \right) \Rightarrow
	\psi
	\left[\begin{array}{c}
		s_i \doteq d \wedge s_j \doteq c \\
		\vee \\
		s_i \doteq s_j \doteq c \wedge m_j \doteq m_i p \\
		\vee \\
		s_i \doteq s_j \doteq d \wedge m_i \doteq m_j p
	\end{array} \right]
	\end{array}
\end{equation*}
for fresh $p \in b^+$.

\item Apply the following rule exhaustively to remove sums: \\
\begin{equation*}
	\begin{array}{c}
	\psi[x_k = x_i + x_j] \\
	\hline
	\left(\begin{array}{l}
		m_i \doteq m_j  \\
		\vee\; m_i \doteq m_j p \\
		\vee\; m_j \doteq m_i p
	\end{array} \right) \Rightarrow
	\psi
	\left[\begin{array}{c}
		s_i \doteq s_j \wedge x_k \doteq a s_i m_i m_j a  \\
		\vee \\
		s_i \not\doteq s_j \wedge m_i \doteq m_j \wedge x_k \doteq aca \\
		\vee \\
		s_i \not\doteq s_j \wedge m_i \doteq m_j p \wedge x_k \doteq a s_i pa \\
		\vee \\
		s_i \not\doteq s_j \wedge m_j \doteq m_i p \wedge x_k \doteq a s_j pa
	\end{array} \right]
	\end{array}
\end{equation*}
for fresh $p \in b^+$.

\item Modify the meaning of regularity constraints as follows: let $\underline{\reg}_i$ be defined by a regular expression with constants in $\integers$.
	Substitute every occurrence of a nonnegative constant $k \in \integers$ by $acb^ka$; every occurrence of a negative constant $k \in \integers$ by $adb^{-k}a$; every occurrence of set $\integers$ by $acb^*a \cup adb^+a$.

\end{enumerate}

The resulting formula is again in form (\ref{eq:prenex}) where $\psi$ is now a quantifier-free formula in the theory of concatenation over $\{a,b,c,d\}^*$; its validity is decidable by Proposition \ref{prop:decidable}.

\subsubsection{Correctness and Complexity}
\label{sec:correctness}
Let us sketch the correctness argument for the decision procedure $\decp$, which shows that the transformed formula is equi-satisfiable with the original one.

The justification for step 1 is straightforward.
After applying it a series of substitutions eliminates arithmetic by reducing it to equations over the theory of concatenation with the unary encoding of integers defined above.

Step 2 requires that any variable $x_i$ is a sequence of the form $(acb^*a \cup adb^+a)^*$ and introduces fresh variables to denote significant parts of the sequence: $h_i$ aliases the first element of the sequence which is further split into its sign $s_i$ ($c$ for nonnegative and $d$ for negative) and its absolute value $m_i$ encoded as a unary string in $b^*$.
The second term of the disjunction deals with the case of $x_i$ being $\epsilon$, which has the same encoding as $0$.

The following steps replace elements of the signature of Presburger arithmetic by rewriting them as equations over sequences with the given encoding.
Step 3 reduces the arithmetic equality of two sequences of integers to equivalence of the sequences encoding their first elements.
Step 4 rewrites equations involving differences with equations involving sums.

Step 5 reduces arithmetic comparisons of two sequences of integers to a case discussion over the sequences $h_i, h_j$ encoding their first elements.
Let $p$ be a sequence in $b^+$ encoding the difference between the absolute values corresponding to $h_i$ and $h_j$; obviously such a $p$ always exists unless the absolute values are equal.
Then, $h_i$ encodes an integer strictly less than $h_j$ iff one of the following holds: (1) $h_i$ is a negative value and $h_j$ is a nonnegative one; (2) both $h_i$ and $h_j$ are a nonnegative value and the sequence of $b$'s in $h_j$ is longer than the sequence of $b$'s in $h_i$; or (3) both $h_i$ and $h_j$ are a negative value and the sequence of $b$'s in $h_i$ is longer than the sequence of $b$'s in $h_j$.

Step 6 reduces the comparison between the value of a sum of two variables and a third variable to an analysis of the three sequences $h_i, h_j, h_k$ encoding the first elements of the three variables.
As in step 6, the unary sequence $p$ encodes the difference between the absolute values corresponding to $h_i$ and $h_j$.
Then, $h_k$ encodes the sum of the values encoded by $h_i$ and $h_j$ iff one of the following holds: (1) $h_i$ and $h_j$ have the same sign and $h_k$ contains a sequence of $b$'s which adds up the sequences of $b$'s of $h_i$ and $h_j$, still with the same sign; (2) $h_i$ and $h_j$ have opposite sign but same absolute value, so $h_k$ must encode $0$; (3) $h_i$ and $h_j$ have opposite sign and the absolute value of $h_i$ is greater than the absolute value of $h_j$, so $h_k$ has the same sign as $h_i$ and the difference of absolute values as its absolute value; or (4) $h_i$ and $h_j$ have opposite sign and the absolute value of $h_j$ is greater than the absolute value of $h_i$, so $h_k$ has the same sign as $h_j$ and the difference of absolute values as its absolute value.

Finally, step 7 details how to translate the interpretation of the regular constraints over $\integers$ into the corresponding regularity constraints over $\{a,b,c,d\}$ with the given integer encoding.

It is not difficult to see that all rewriting steps in the decision procedure $\decp$ increase the size of $\psi$ at most quadratically (this accounts for fresh variables as well).
Hence, the PSPACE complexity of the universal fragment of the theory of concatenation (Proposition \ref{prop:decidable}) carries over to $\decp$.

\begin{theorem}
The universal fragment of $\seqth$ is decidable in PSPACE with the decision procedure $\decp$.
\end{theorem}

\subsection{Undecidable Extensions}
\label{sec:extensions}

\begin{theorem}
The following extensions of the universal fragment of $\seqth$ are undecidable.
\begin{enumerate}
\item The $\forall^* \exists^*$ and $\exists^* \forall^*$ fragments.

\item For any pair of integer constants $k_1, k_2$, the extension with the two length functions $|x|_{k_1}, |x|_{k_2}$ counting the number of occurrences of $k_1$ and $k_2$ in $x$.

\item The extension with an equal length predicate $Elg(x,y) \triangleq |x| = |y|$.

\item The extension with a sum function $\sigma(x) \triangleq \sum_{i = 1}^{|x|} x(i,i)$.
\end{enumerate}
\end{theorem}

\begin{proof}
\begin{enumerate}
\item
Sentences with one quantifier alternation are undecidable already for the theory of concatenation without arithmetic and over a monoid of finite rank (Proposition \ref{prop:undecidable}).
Notice that the set of sentences that are expressible both in the $\forall^* \exists^*$ and in the $\exists^* \forall^*$ fragment is decidable \cite[Th.~4.4]{Sei92}; however, this set lacks a simple syntactic characterization.

\item
Corollary of Proposition \ref{prop:undecidable}.

\item
We encode the universal theory of $\Pi = \langle \naturals, 0, 1, +, \pi \rangle$ --- where $\pi(x,y) \triangleq x 2^y$ --- in the universal fragment of $\seqth$ extended by the $Elg$ predicate; undecidability follows from the undecidability of the existential and universal theories of $\Pi$ \cite[Corollary 5]{BS-definability}.
All we have to do is showing that $\pi(x,y) = p$ is universally definable in $\seqth$ with $Elg$.
To this end, first define $l_y$ as a sequence that begins with value $y$, ends with value $1$, and where every element is the successor of the element that follows.
\begin{equation*}
\forall h,t \scope
\first(l_y) = y  \wedge \last(l_y) = 1 \wedge l_y \doteq ht \wedge |h| > 0 \wedge |t| > 0  \Rightarrow \last(h) = t + 1
\end{equation*}
As a result $l_y$ is in the form $y , y-1, \ldots, 1$ and hence has length $y$.\footnote{This technique would allow the definition of the length function $|x|$ and full index arithmetic as well.}
Then, $\pi(x,y)$ is universally definable as the sequence $p$ with the same length as $l_y$, whose last element is $x$, and where every element is obtained by doubling the value of the element that follows:
\begin{equation*}
\forall g,u \scope
Elg(p, l_y) \wedge \last(p) = x \wedge p \doteq gu \wedge |g| > 0 \wedge |u| > 0 \Rightarrow \last(g) = u + u
\end{equation*}
Hence $p$ has the form $2^y x, 2^{y-1}x, \ldots, 2^2x, 2x, x$ which encodes the desired value $x 2^y$ in $\seqth$.
(Notice that the two universal definitions of $l_y$ and $p$ can be combined into a single universal definition by conjoining the definition of $p$ to the consequent in the definition of $l_y$).

\item
For any sequence $x$ over $\{0,1\}$ define $Sp(x) = y$ as $y \in 0^*1^* \wedge \sigma(y) = \sigma(x)$.
Then, Proposition \ref{prop:undecidable} implies undecidability because this extension of $\seqth$ can define universal sentences over the free monoid $\{a,b\}^*$ with the function $Sp$. \qedhere

\end{enumerate}
\end{proof}

The decidability of the following is instead currently unknown: the extension of the universal fragment with a function $x \oplus 1$ defined as the sequence $x(1)+1, x(2)+1, \ldots, x(|x|)+1$.
The fragment allows the definition of the set $S^= \{0^n 1^n \mid n \in \naturals \}$ as the sequences $x$ such that $x \in 0^* 1^* \wedge \forall u,v \scope x \doteq u v \wedge u \in 0^* \wedge v \in 1^* \Rightarrow u \oplus 1 \doteq v$.
This is inexpressible in the universal fragment of the theory of concatenation, but the decidability of the resulting fragment is currently unknown (see Proposition \ref{prop:definable}).

\section{Verifying Sequence-Manipulating Programs}
\label{sec:programs}
This section outlines a couple of examples that demonstrate using formulas in the theory $\seqth$ to reason about sequence-manipulating programs.
An implementation of the decision procedure $\decp$ is needed to tackle more extensive examples; it is currently underway.
The examples are in Eiffel-like pseudo-code \cite{OOSC2}; it is not difficult to detail an axiomatic semantics and a backward substitution calculus, using the universal fragment of $\seqth$, for the portions of this language used in the examples.

\begin{table}[!tb]
\begin{scriptsize}
\centering
\begin{tabular}{m{.65\textwidth} m{.35\textwidth}}
\begin{lstlisting}
merge_sort (a: ARRAY): ARRAY
local l,r: ARRAY
do
   if |a| $\leq$ 1 then
      { sorted(a) }
      Result := a
   else
      l , r := a[1:|a|/2] , a[|a|/2+1: |a|]
      { l $\cat$ r = a }
      l , r := merge_sort (l) , merge_sort (r)
      { sorted(l) $\wedge$ sorted(r) }
      from Result := $\epsilon$
      { invariant sorted(Result) $\wedge$ sorted(l) $\wedge$ sorted(r) $\wedge$
                   $\last$(Result) $\leq$ $\first$(l) $\wedge$ $\last$(Result) $\leq$ $\first$(r) }
      until |l| = 0 $\vee$ |r| = 0
      loop
         if l.first > r.first then
           Result := Result $\cat$ r.first ; r := r.rest
         else
           Result := Result $\cat$ l.first ; l := l.rest
         end
      end
      if |l| > 0 then
         { |r| = 0 }  Result := Result $\cat$ l
      else
         { |l| = 0 }  Result := Result $\cat$ r
      end
{ ensure  sorted(Result) }
\end{lstlisting}
&
\begin{lstlisting}
reverse (a: LIST): LIST
local v: INTEGER ; s: STACK
do
   from s := $\epsilon$
   { invariant s $\catl$ a = old a } 
   until a = $\epsilon$
   loop
      s.push (a.first)
      a := a.rest
   end
   from Result := $\epsilon$
   { invariant
        s $\catl$ Result$\rev$ = old a }
   until s = $\epsilon$
   loop
      v := s.top
      s.pop ; Result.extend (v)
   end
{ ensure Result$\rev$ = old a}
\end{lstlisting}
\end{tabular}
\end{scriptsize}
\caption{Annotated Mergesort (left) and Array Reversal (right).}
\label{tab:merge-sort}
\end{table}

\paragraph{Reversal.}
In Table \ref{tab:merge-sort} (right), a program reverses a sequence of integers, given as a list $a$, using a stack $s$.
The query ``first'' returns the first element in a list, and the command ``extend'' adds an element to the right of a list; the query ``top'' and the commands ``pop'' and ``push'' for a stack have the usual semantics.
In the annotations, $s$ is modeled by a sequence whose first element is the bottom of the stack, whereas the expression $\old\;a$ denotes the value of $a$ upon entering the routine.

The superscript $\rev$ denotes the reversal of a sequence.
We do not know if the extension of $\seqth$ by a reversal function is decidable.
However, the following two simple update axioms are sufficient to handle any program which builds the reverse $\uu\rev$ of a sequence $\uu$ starting from an empty sequence and adding one element at a time:
\begin{align*}
 \uu\rev = \epsilon &\ \Leftrightarrow\ \uu = \epsilon
&
|x| = 1  &\ \Rightarrow\ (\uu x)\rev = x \uu\rev
\end{align*}

Consider, for instance, the verification condition that checks if the invariant of the second loop (lines 11--18) is indeed inductive:
\begin{equation*}
s \catl \Result\rev = \old\;a
\wedge s \neq \epsilon
\quad \Rightarrow \quad
s(1,-1) \catl (\Result \catl s(0,0))\rev = \old\;a
\end{equation*}
After rewriting $(\Result \catl s(0,0))\rev$ into $s(0,0) \catl \Result\rev$ the implication is straightforward to check for validity.
The rest of the program is also simple to check with standard backward reasoning techniques.

\paragraph{Mergesort.}
Consider a straightforward recursive implementation of the Mergesort algorithm; Table \ref{tab:merge-sort} (left) shows an annotated version, where $\cat$ denotes the concatenation operator in the programming language (whose semantics is captured by the corresponding logic operator $\catl$).
The annotations specify that the routine produces a sorted array, where predicate $\sorted(\uu)$ is defined as:
\begin{equation*}
\sorted(\uu) \quad\triangleq\quad \forall h,m,t \scope \uu \doteq hmt \wedge |m| > 1 \wedge |t| > 0 \Rightarrow m \leq t
\end{equation*}
It is impossible to express in $\seqth$ another component of the full functional specification: the output is a permutation of the input.
This condition is inexpressible in most of the expressive decidable extensions of the theory of arrays that are currently known, such as \cite{BMS06-vmcai,HIV08-lpar} (see also Section \ref{sec:related}).
Complementary automated verification techniques --- using different abstractions such as the multiset \cite{PK-multiset} --- can, however, verify this orthogonal aspect.

We must also abstract away the precise splitting of array $a$ into two halves in line 8.
The way in which $a$ is partitioned into $l$ and $r$ is however irrelevant as far as correctness if concerned (it only influences the complexity of the algorithm), hence we can simply over-approximate the instruction on line 8 by a nonderministic splitting in two continuous non-empty parts.

From the annotated program, we can generate verification conditions by standard backward reasoning.
Universal sentences of $\seqth$ can express the verification conditions, hence the verification process can be automated.
Let us see an example on the non-trivial part of the process, namely checking that the formula on lines 13--14 is indeed an inductive invariant.
Consider the ``then'' branch on line 18.
Backward substitution of the invariant yields:
\begin{gather}
\sorted(\Result \ast \first(r)) \;\wedge\; \sorted(l) \;\wedge\; \sorted(r(2,0)) \;\wedge\; \nonumber \\
\last(\Result \ast \first(r)) \leq \first(l) \;\wedge\; \last(\Result \ast \first(r)) \leq \first(r(2,0))  \label{eq:merge-bsub}
\end{gather}

This condition must be discharged by the corresponding loop invariant hypothesis:
\begin{gather}
\first(l) > \first(r) \;\wedge\; \sorted(\Result) \;\wedge\; \sorted(l) \;\wedge\; \sorted(r) \;\wedge\; \label{eq:merge-hyp}  \\
\last(\Result) \leq \first(l) \;\wedge\; \last(\Result) \leq \first(r) \;\wedge\;
|l| \neq 0  \;\wedge\; |r| \neq 0   \nonumber
\end{gather}
Checking that (\ref{eq:merge-hyp}) entails (\ref{eq:merge-bsub}) discharges the corresponding verification condition.
Elements of this condition can be encoded in the universal fragment of $\seqth$ and proven using the decision procedure of Section~\ref{sec:decision}; for instance, the fact that $\last(\Result) \leq \first(l)$, $|l| \neq 0$, $|r| \neq 0$, and $\first(l) > \first(r)$ imply $\last(\Result \ast \first(r)) \leq \first(l)$ corresponds to the validity of (all free variables are implicitly universally quantified):
\begin{equation*}
\left( \begin{array}{l}
    r \doteq h_r m_r t_r  
    \ \wedge\  |h_r| = 1 \ \wedge\ |r| \neq 0 \\
    \wedge\;\; l \doteq h_l m_l t_l 
    \ \wedge\  |h_l| = 1 \ \wedge\ |l| \neq 0 \\
    \wedge\;\;\Result \catl h_r \doteq hmt \wedge |t| = 1 \\
    \wedge\;\; h_l > h_r
\end{array} \right)
 \Rightarrow
 t \leq h_l
\end{equation*}


\section{Related Work}
\label{sec:related}
Pioneering efforts on automated program verification focused on very simple data types --- in most cases just scalar variables --- as the inherent difficulties were already egregious.
As verification techniques progressed and matured, more complex data types were considered, such as lists (usually \emph{\`a la} Lisp), arrays, maps, and pointers, up to complex dynamic data structures.
Arrays in particular received a lot of attention, both for historical reasons (programming languages have been offering them natively for decades), and because they often serve as the basis for implementing more complex data structures.
More generally, a renewed interest in developing decision procedures for new theories and in integrating existing ones has blossomed over the last few years.
A review of this staggering amount of work is beyond the scope of this paper; for a partial account and further references we refer the reader to e.g., \cite{ZKR08,KPSW10} (and \cite{HW09,KGGHE09} for applications).
In this section, we review approaches that are most similar to ours and in particular which yield decidable logics that can be compared directly to our theory of sequences (see Section \ref{sec:examples}).
This is the case with several of the works on the theory of arrays and extensions thereof.

\paragraph{The theory of arrays.}
McCarthy initiated the research on formal reasoning about arrays \cite{McC62}.
His theory of arrays defines the axiomatization of the basic access operations of \emph{read} and \emph{write} for quantifier-free formulas and without arithmetic or extensionality (i.e., the property that if all elements of two arrays are equal then the arrays themselves are equal).
McCarthy's work has usually been the kernel of every theory of arrays: most works on (automated) reasoning about arrays extend McCarthy's theory with more complex (decidable) properties or efficiently automate reasoning within an existing theory.

Thus, a series of work extended the theory of arrays with arithmetic \cite{SJ80-jacm,Jaf81-ipl} and with sorting predicates on array segments \cite{Mat81-jacm}; automated reasoning within these theories is possible only for restricted classes of programs.
Extensionality is another very significant extension to the theory of arrays \cite{SBDL01-lics}, which has now become standard as it is decidable.

The fast technological advances in automated theorem proving over the last years have paved the way for efficient implementations of the theory of arrays (usually with extensionality).
These implementations use a variety of techniques such as SMT solving \cite{ARR01-csl,GD07,BNOCR08,GNRZ08-ijcar,GKF08}, saturation theorem proving \cite{LM02,ABRS09}, and abstraction \cite{BB08,JMcM07-cav,KMZ06}.
Automated invariant inference is an important application of these decision procedures, which originated a specialized line of work \cite{BHMR07-vmcai,McM08-tacas,KV09-fase}.

\paragraph{Decidable extensions of the theory of arrays.}
The last few years have seen an acceleration in the development of decidable extensions of the extensional theory of arrays with more expressive predicates and functions.

Bradley et al.\ \cite{BMS06-vmcai} develop the \emph{array property fragment}, a decidable subset of the $\exists^* \forall^*$ fragment of the theory of arrays.
An \emph{array property} is a formula of the form $\exists^* \forall^* \scope \iota \Rightarrow \nu$, where the universal quantification is restricted to index variables, $\iota$ is a guard on index variables with arithmetic (restricted to existentially quantified variables), and $\nu$ is a constraint on array values without arithmetic or nested reads, and where no universally quantified index variable is used to select an element that is written to.
The array property fragment is decidable with a decision procedure that eliminates universal quantification on index variables by reducing it to conjunctions on a suitable finite set of index values.
Extensions of the array property fragment that relax any of the restrictions on the form of array properties are undecidable.
Bradley et al.\ also show how to adapt their theory of arrays to reason about maps.

Ghilardi et al.\ \cite{GNRZ07-amai} develop ``semantic'' techniques to integrate decision procedures into a decidable extension of the theory of arrays.
Their $\adp$ theory merges the quantifier-free extensional theory of arrays with dimension and Presburger arithmetic over indices into a decidable logic.
Two extensions of the $\adp$ theory are still decidable: one with a unary predicate that determines if an array is \emph{injective} (i.e., it has no repeated elements); and one with a function that returns the \emph{domain} of an array (i.e., the set of indices that correspond to definite values).
Ghilardi et al.\ suggest that these extensions might be the basis for automated reasoning on Separation Logic models.
The framework of \cite{GNRZ07-amai} also supports other decidable extensions, such as the \emph{prefix}, and \emph{sorting} predicates, as well as the \emph{map} combinator also discussed in \cite{dMB09}.

De Moura and Bj{\o}rner \cite{dMB09} introduce \emph{combinatory array logic}, a decidable extension of the quantifier-free extensional theory of arrays with the \emph{map} and \emph{constant-value} combinators (i.e., array functors).
The \emph{constant-value} combinator defines an array with all values equal to a constant; the \emph{map} combinator applies a $k$-ary function to the elements at position $i$ in $k$ arrays $a_1, \ldots, a_k$.
De Moura and Bj{\o}rner define a decision procedure for their combinatory array logic, which is implemented in the Z3 SMT solver.

Habermehl et al.\ introduce powerful logics to reason about arrays with integer values \cite{HIV08-fossacs,HIV08-lpar,BHIKV09-cav}; unlike most related work, the decidability of their logic relies on automata-theoretic techniques for a special class of counter automata.
More precisely, \cite{HIV08-fossacs} defines the \emph{Logic of Integer Arrays} \lia{}, whose formulas are in the $\exists^* \forall^*$ fragment and allow Presburger arithmetic on existentially quantified variables, difference and modulo constraints on index variables, and difference constraints on array values.
Forbidding disjunctions of difference constraints on array values is necessary to ensure decidability.
The resulting fragment is quite expressive, and in particular it includes practically useful formulas that are inexpressible in other decidable expressive fragments such as \cite{BMS06-vmcai}.
The companion work \cite{HIV08-lpar} introduces the \emph{Single Index Logic} \sil{}, consisting of existentially quantified Boolean combinations of formulas of the form $\forall^* \scope \iota \Rightarrow \nu$, where the universal quantification is restricted to index variables, $\iota$ is a positive Boolean combination of bound and modulo constraints on index variables, and $\nu$ is a conjunction of difference constraints on array values.
Again, the restrictions on quantifier alternations and Boolean combinations are tight in that relaxing one of them leads to undecidability.
The expressiveness of \sil{} is very close to that of \lia{}, and the two logics can be considered two variants of the same basic kernel.
The other work \cite{BHIKV09-cav} shows how to use \sil{} to annotate and reason automatically about array-manipulating programs; the tight correspondence between \sil{} and a class of counter automata allows the automatic generation of loop invariants and hence the automation of the full verification process.

\paragraph{Other approaches.}
Static analysis and abstract interpretation techniques have also been successfully applied to the analysis of array operations, especially  with the goal of inferring invariants automatically (e.g., \cite{GRS05-popl,GMT08-popl,HP08-pldi}).

\section{Future Work}
\label{sec:conclusion}
Future work will investigate the decidability of the universal fragment of $\seqth$ extended with ``weak'' predicates or functions that slightly increase its expressiveness (such as that outlined at the end of Section \ref{sec:extensions}).
We will study to what extent the decision procedure for the universal fragment of $\seqth$ can be integrated with other decidable logic fragments (and possibly with the dual existential fragment).
We will investigate how to automate the generation of inductive invariants for sequence-manipulating programs by leveraging the decidability of the universal fragment of $\seqth$.
Finally, we will implement the decision procedure, integrate it within a verification environment, and assess its empirical effectiveness on real programs.


\end{document}